\documentclass[grl]{AGUTeX}
\usepackage{color}
\usepackage{amsmath}
\usepackage{amsfonts}
\usepackage{graphicx}
\authorrunninghead{HEYMAN et al.}
\titlerunninghead{BEDLOAD, TIME SCALES AND COLL. MOTION}

\authoraddr{Joris Heyman,
LHE Station 18, EPFL, Lausanne, 1015, Switzerland.
(joris.heyman@epfl.ch)}
\authoraddr{HongBo Ma, State Key Laboratory of Hydroscience and Engineering, Department of Hydraulic Engineering, Tsinghua University, Beijing 100084, China.
(mhb05@mails.tsinghua.edu.cn)}

\begin{document}

\title{Statistics of bedload transport over steep slopes: Separation of time scales and collective motion.}

 \authors{J. Heyman,\altaffilmark{1}
 F. Mettra,\altaffilmark{1} H.B. Ma,\altaffilmark{1,}\altaffilmark{2}
 C. Ancey \altaffilmark{1}}
\altaffiltext{1}{Laboratory of Environmental Hydraulics,  School of Architecture, Civil and Environmental Engineering, \'Ecole Polytechnique F\'ed\'erale de Lausanne, Switzerland}
\altaffiltext{2}{State Key Laboratory of Hydroscience and Engineering, Tsinghua University, Beijing, China}

\begin{abstract}
Steep slope streams show large fluctuations of sediment discharge across several time scales. These fluctuations may be inherent to the internal dynamics of the sediment transport process.  A probabilistic framework thus  seems appropriate to analyze such a process. In this letter, we present an experimental study of bedload transport over a steep slope flume for small to moderate Shields numbers. 
The sampling technique allows the acquisition of high-resolution time series of the solid discharge. The resolved time scales range from $10^{-2}$s up to $10^{5}$s. We show that two distinct time scales can be observed in the probability density function for the waiting time between moving particles. We make the point that the separation of time scales is related to collective dynamics. Proper statistics of a Markov process including collective entrainment are derived. The separation of time scales is recovered theoretically for low entrainment rates.
\end{abstract}

\begin{article}
\section{Introduction}
Steep slope streams carry large amounts of sediment from glaciers down to  valleys. As the mean diameter of grains is relatively large in these rivers, sediment transport mainly takes the form of bedload: particles are never carried entirely by the fluid but slide, roll and move downstream by small jumps (saltation). 

Despite being a key process in mountainous landscape evolution, bedload transport in steep slope rivers is still poorly understood and largely unpredictable. For instance, prediction from modern formulae may diverge by between 1 and 4 orders of magnitude from field data, leaving little hope of quantifying any cumulative sediment volume at long times. 

There exist several reasons for the failure of traditional equations when applied to steep slope streams, the most important one being the complexity of the flow in such rivers. For instance, large protruding boulders significantly reduce the transport capacity of the streams and thus often lead to over-prediction of solid discharge \citep{Yager2012}. On the other hand, full transport capacity  is also limited by sediment availability, which in turn is difficult to estimate in the field.

Furthermore, the dynamics of moving particles on steep slopes are known to be highly nonlinear and  intermittent. The particle flow rate may exhibit chaotic behavior at certain scales. Early experimental results have already pinpointed the fluctuating behavior of bedload discharge, even in precisely controlled flows and under idealized conditions \citep{kuhnle}.  Recent results show that steep slope channels are particularly prone to such variability \citep{ANCEY2006,Zimmermann2010} and strongly suggest the use of a probabilistic framework to model bedload transport. 

The first attempt to derive a statistical equation is attributed to \citet{Einstein}. Analyzing the random motion of particles, Einstein found that bedload discharge should follow a Poisson distribution. Since then, many probabilistic equations have been derived for erosion/deposition models \citep{Lajeunesse2010}. \citet{turowski2010} derived the probability density function (pdf) of the cumulative solid discharge assuming that the distribution of the waiting time between two moving particles is known and that all particles move independently. The Poisson distribution is recovered when waiting times are exponentially distributed. However, Poissonian models seem to under-predict the variability of bedload transport over steep slopes \citep{ANCEY2006}.

Recently, many studies have shown the singular fractal and intermittent characteristics of bedload data series, and their probable origin in long range correlated processes \citep{singh2009}. \citet{ANCEY2008} proposed a Markovian model based on a reduced set of possible particle movements  and showed that large fluctuations around the mean and correlation were possible. Based on their experimental observations,  \citet{ANCEY2008} introduced a collective entrainment parameter to account for the collective motion of particles. 

Here, we present the results of an experimental study of bedload transport in a steep laboratory flume. We show that there exists a net separation between time scales in the statistics of the solid discharge. Moreover, we demonstrate that the Markovian model of \citet{ANCEY2008} also predicts a separation between time scales when collective dynamics are considered and for flow close to the onset of particle motion.  Thus, we conclude that collective dynamics are of particular relevance in bedload transport over steep slopes at incipient motion conditions.

\section{Collective motion}
Collective dynamics have often been introduced into continuous models of dry granular avalanches on slopes close to the critical angle of stability \citep{bouchaud1995}. Regarding granular media sheared by air, collective entrainment of particles occurs frequently when saltating grains impact the bed and eject other ones. According to \citet{bagnoldbook} this should not be the case in water, where the density difference between the particles and the fluid is much lower. However, collective motion of bedload particles in water has been reported in-situ by \citet{Drake1988,Dinehart1999} while \citet{ANCEY2008} noticed it on a steep laboratory flume. Several explanations can be given:
\begin{enumerate}
\item The critical angle of stability for grains sheared by a fluid is known to be significantly reduced \citep{loiseleux}. As avalanche precursors might happen below the critical angle of stability \citep{staron}, small avalanches, triggered by the entrainment of a particle, are likely to propagate locally.
\item When the mean shear velocity is close to the threshold of incipient motion, small disturbances may initiate particle motion. This is true for turbulent eddies that locally dislodge particles \citep{papa2002}. Fast-moving particles that settle on the bed may also potentially disturb resting particles. Particle momentum grows as $d^3$ while drag force as $d^2$ so that beyond a critical diameter, particle-particle interactions should become more important than fluid-particle interactions. As steep slope streams generally convey large particles, particle-particle interactions may be an important mode of entrainment.
\item Turbulent eddies are spatially correlated, and thus may dislodge several particles simultaneously, leading to  ``clouds'' of moving particles. \citet{Drake1988} described them as  "sweep-transport" events, that occur only about 9\% of the time but still account for 90\% of the cumulative load, highlighting thus their importance. 
\item More generally, any kind of time correlation in stationary time series (such as bedform migration), if short enough, can in principle be modelled by a collective feedback to the equations.
\end{enumerate} 
Although the relative importance of each of these phenomena in the collective entrainment of bed particles is not fully understood today, they all have the same eventual effect on the temporal dynamics of bedload (intermittent periods of no transport, separated by intense bursts). The simplest theoretical approach thus consists of merging all possible physical causes into a generic collective feedback mechanism.
\section{Experiments}
\begin{table}
\caption{Experimental conditions.}
\centering
\begin{tabular}{r|cccccccccc}
  &$\tau_s$   &Fr & $\theta$ & $\bar{u}$ & $\bar{h}$ &$\bar{q_s}$&$\bar{T}$&$\mbox{Var}[T]/\bar{T}$\\
    \hline
   (a)&0.079   & 1.44 &  7 & 0.53& 1.37 & 1.11 & 0.89&5.49\\
      \hline
   (b)&0.088  & 1.48 &  7 &0.60 & 1.68& 5.87  & 0.17&1.53 \\
      \hline
   (c)&0.092 &1.50 &  7 &0.62&1.75 & 13.38&0.078&0.24\\
      \hline
\end{tabular}
\label{table1}
\tablenotetext{}{$\tau_s$, Shields stress; Fr, Froud number; $\theta$, slope angle (\%); $\bar{u}$, mean fluid velocity (m/s); $\bar{h}$, mean water depth (cm); $\bar{q_s}$, mean output solid discharge (particles/s); $\bar{T}$, mean waiting time between emigration events (s).}
\end{table}

\begin{figure*}
\includegraphics[width=40pc]{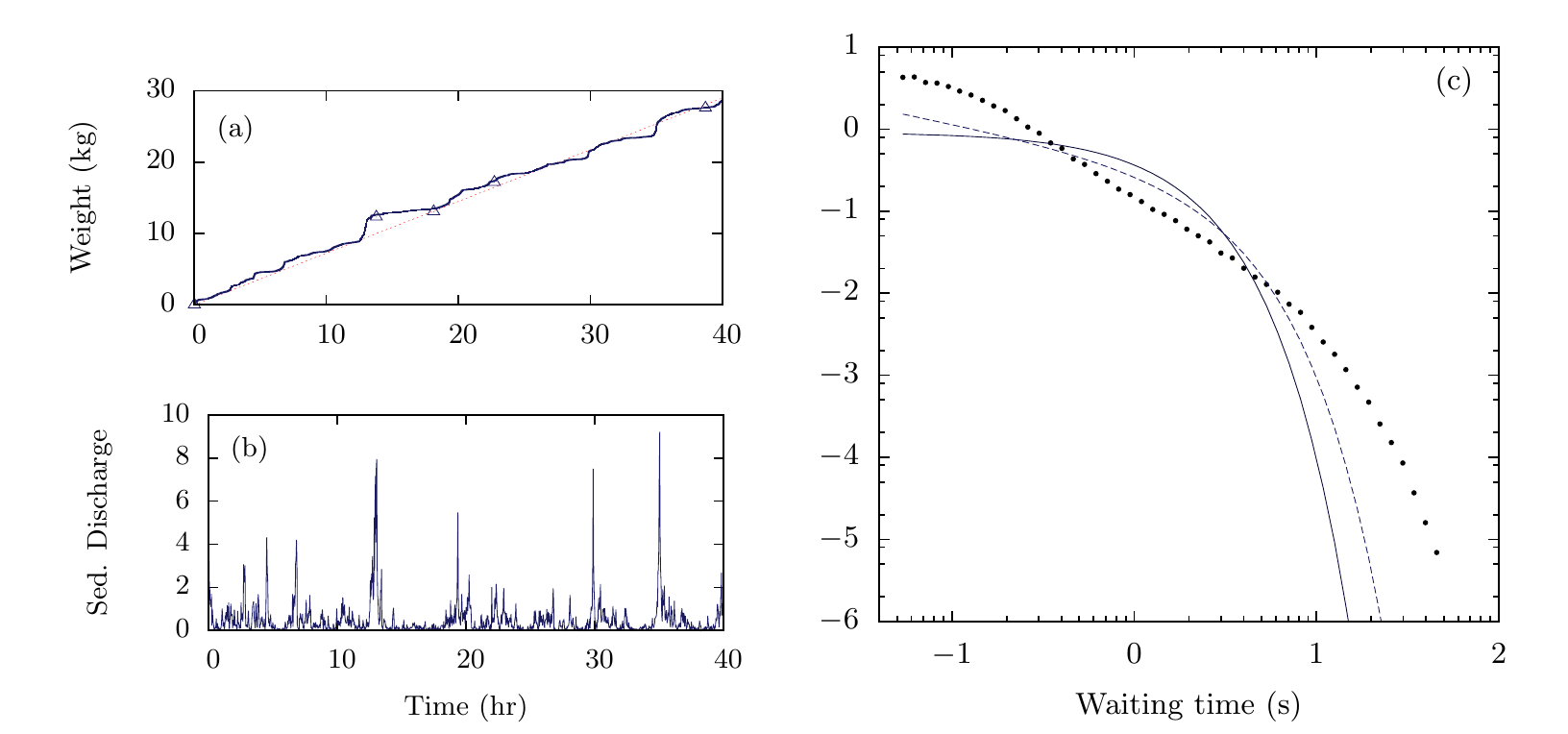}
\caption{Results of experiment (a) at low Shields stress. (a) Cumulative bedload weight measured by accelerometers: Input (dashed line) and output (line).  Direct weightings of output (triangles). (b) Mean output solid discharge (particles/s). (c) Pdf (log-log) for the waiting time between moving particles (dots), Exponential fit (line) and Gamma fit (dashed line).}
\label{figure1}
\end{figure*}

We conducted 48-hour experiments on a 2.5 meter long steep slope flume. The erodible bed was made of uni-sized natural sediment particles of mean diameter 8$\pm$1.5 mm. The flume was 8 cm wide and the water depth ranged from 1 to 3 cm during experiments. This setup was able to reproduce the fully turbulent ($Re\ge 8000$) and supercritical flow of natural steep rivers, while avoiding complex 3-dimensional patterns like meanders and bars owing to the high water depth to channel width ratio. Moreover, the extremely narrow grain size distribution limited fluctuations due to size segregation and other related phenomena \citep{freychurch2009}. 

Three experiments of increasing Shields stress, called (a), (b) and (c) are presented here. In each experiment, we kept the slope constant and adjusted the water discharge to reach the intended bed shear stress. In addition, the sediment input was controlled to guarantee a global equilibrium between erosion and deposition and to maintain the mean slope value. Adimensional numbers and experimental parameters are given in Table \ref{table1}.
As a consequence of the fairly high Froude number ($\mbox{Fr}\sim 1.4$), antidunes, scaling with the water depth, formed in all experiments and propagated upstream along the channel. Their lee face angle ranged from 10 to $20^\circ$, close to the angle of repose of a slope sheared by a fluid \citep{loiseleux}. Entrainment of particles in those regions was highly intermittent, as can be seen in the video available online.

Sediment fluxes were measured both at the entrance and at the output of the flume by a new technique detailed as follows. Upon leaving the channel, each particle hits a metallic plate and the impact is recorded by a small accelerometer tied to the plate. A peak-over-threshold method is then applied to detect the times of the consecutive particle impacts. If sufficient damping is provided to avoid spurious vibrations of the plate, this method ensures an extremely good time resolution in bedload discharge rates. As the grain size distribution is narrow, little error is made when converting the number of particles to the sediment weight. Finally, the time series covered at least 6 orders of magnitude (from $\sim 10^{-1}$ to $10^{5}$ s). To validate the method, we measured the output weight of sediment systematically during experiments (Fig.~\ref{figure1}a). 

For clarity, we only present the results of experiment (a) in Fig.~\ref{figure1}. Results of (b) and (c) will be compared  to (a) in the last section. As expected, bedload discharge shows great variability over a wide range of time scales (Fig.~\ref{figure1}b). In this letter, we focus on the statistics of the waiting time between two particles leaving the channel. This variable (denoted by $T$) is of particular relevance to sediment transport, because its inverse gives a measure of the instantaneous bedload discharge. The chosen measurement technique allows us to obtain the waiting time directly by differentiating the time series of outgoing particles (see equation $\eqref{diff}$) . The pdf of $T$ is then estimated from more than $10^5$ individual waiting time samples.

The first striking feature of the pdf plotted in Fig.~\ref{figure1}c is that its tail is much thicker than the exponential distribution, which is commonly used in probabilistic bedload models \citep{Einstein,turowski2010}. The gamma distribution also poorly represents the pdf for the waiting time. A closer look at Fig.~\ref{figure1}c shows an even more interesting feature: two bumps can be distinguished on the curve, a fast mode (short times) and a slow mode (long times). One can interpret this bimodal shape by the coexistence of two processes that act on very different time scales.

We show in the following that this bimodal distribution is likely to be generated by the collective motion of particles. Moreover, we will show that a simple Markov model is able to mimic this distribution, despite being memoryless (the future of the process only depends on its present state, so that the waiting time between state change is exponentially distributed \citep{Gillespie1977}). 

\section{Theory}
\subsection{The Continuous Time Markov Process}

\begin{figure}[t]
\centering
\includegraphics[width=20pc]{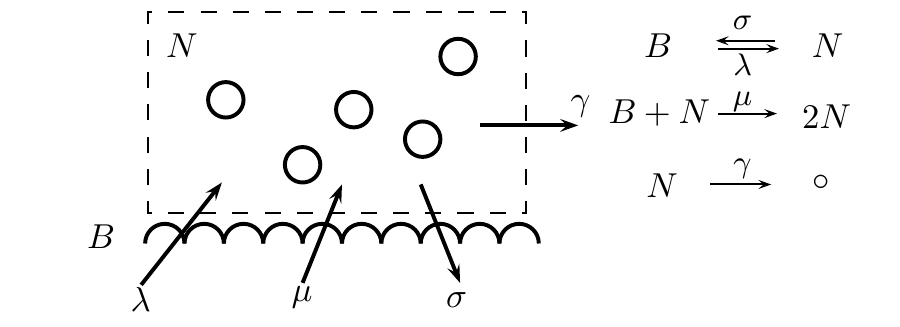}
\caption{Sketch of the model and corresponding reactions. $N$ is the number of moving particles while $B$ is the number of particles in the bed.}
\label{figure2}
\end{figure}

Our starting point is the Markov model of \citet{ANCEY2008}, that considers the number of moving particles in an observation window at time $t$ ($N(t)\in \mathbb{Z}^+$) to be the main random variable. $N(t)$ evolves in time according to a continuous-time birth-death Markov process (Fig.~\ref{figure2}). 

Entrainment of a particle by the fluid (rate $\lambda$ [particles.$\mbox{s}^{-1}$]), deposition of a moving particle onto the bed (rate $\sigma$ [$\mbox{s}^{-1}$]) and emigration of a moving particle out of the window (rate $\gamma$ [$\mbox{s}^{-1}$]) are all probable events that change the value of $N(t)$. \citet{ANCEY2008} also introduced the collective entrainment phenomena, by the probable event that a moving particle disturbs and ejects another one (with rate $\mu$ [$\mbox{s}^{-1}$]). In experiments, sediment feeding is  adjusted to achieve equilibrium between erosion and deposition and to maintain a constant bed slope. In our model, this is equivalent to assuming that the bed reservoir ($B$) has an infinite capacity so that the process is stationary.

Note that all the parameters have a physical meaning: $\gamma$ is proportional to the particle velocity divided by the window length,  $\sigma$ can be determined independently thanks to imaging techniques, whereas $\lambda$ and $\mu$ are harder to determine by image analysis since entrainment and collective entrainment have the same observable effect (e.g., the entrainment of a particle).  

In \citet{ANCEY2008}, the number of moving particles in an observation window was obtained with a high speed camera. This technique gives direct access to the state of $N(t)$.  However, it bears some disadvantages: (i) only short samples may be acquired due to the large amount of data for storage and treatment and (ii) favourable and controlled conditions are necessary for image processing to be accurate. Its use in natural rivers is thus difficult. Hence, most experimental devices sample the solid flux at only one location (via weighting traps or geo-sensors for instance) through time. This is also the case in this study.

In the following we show that it is still possible to link the model of \citet{ANCEY2008}, defined on an observation window, to the solid discharge obtained at a precise location. In the model, emigration events can be physically associated to particles leaving the channel, or passing through a plane. Thus, determining theoretically the statistics of  those events allows us to compare the experimental solid discharge to the birth-death process.  

Calling $S_k$ the total time when the $k\mbox{-th}$ emigration event occurs, we can define the variable of interest $T_k \in \mathbb{R}^+_*$ such that:

\begin{equation}
S_k=\sum_{i=0}^k T_i
\end{equation}
or equivalently:

\begin{equation}
T_k=S_k-S_{k-1}
\label{diff}
\end{equation}

$T_k$ thus describes the waiting time between emigration events. Assuming that the process is stationary (if $t\rightarrow \infty$ and $\gamma+\sigma>\mu$), $T_k$ are identically distributed and we can drop the $k$ index.  It is important to note that, thanks to the memoryless property of Markovian processes, the waiting time between two jumps of $N$ given the present state is exponentially distributed \citep{Gillespie1977}. However, there are no apparent reasons for the probability of $T$ to be exponentially distributed. 

\subsection{Statistics of $T$}
 Let us define $ F_n$ such that:
 
\begin{equation}
 F_n(t)=\Pr(T>t , N(t)=n),
\end{equation}

where $t$ is the time since the last emigration event. $F_n(t)$ is the probability that $N(t)=n$ and that no emigration event occurred during time $t$. We have:
\begin{equation}
 F(t)=\Pr(T>t)=\sum_{n=0}^{\infty}  F_n(x).
\label{totalX}
\end{equation}
We now consider the transition probabilities on the interval $(t,t+\Delta t]$. $F_n(t+\Delta t)$ is equivalent to the probability that any other event but no emigration occurs during  $\Delta t$. For the state $N(t+\Delta t)=0$, we have thus:
\begin{equation}
  F_0(t+\Delta t)= F_0(t)\left[1-\lambda \Delta t\right] +  F_{1}(t)\sigma \Delta t +o(\Delta t) ,
 	\end{equation}
 	 where the first term on the RHS 	is the probability that nothing happened during $\Delta t$. In the case $N(t)=0$, it is the probability of no entrainment. The second term on the RHS is the probability that a unique moving  particle ($N(t)=1$) deposited onto the bed so that $N(t+\Delta t)=0$. Similarly, for any $n\geq1$:
\begin{eqnarray}
  F_n(t+\Delta t)&=& F_n(t)\left[1-(\lambda+n(\sigma+ \mu+\gamma)) \Delta t\right] \nonumber \\
 	&+&  F_{n+1}(t)\sigma (n+1) \Delta t   \\
 	&+&  F_{n-1}(t)\left[\lambda+\mu (n-1)\right] \Delta t \nonumber \\
 	&+&o(\Delta t) \nonumber.
\end{eqnarray}
Dividing by $\Delta t$ and letting $\Delta t \rightarrow 0$, we get:
\begin{eqnarray}
 F'_0(t)&=&-\lambda  F_0(t) + \sigma  F_{1}(t) \nonumber \\
F'_n(t) &=&-(\lambda+n(\sigma+ \mu+\gamma) F_n(t)  \\
&+& \sigma (n+1)  F_{n+1}(t)  \nonumber \\
&+&\left[\lambda+\mu (n-1)\right] F_{n-1}(t)  \nonumber  \mbox{~ for~} n\geq1,
	\label{masterF}
\end{eqnarray}
where the prime indicates the time derivative. The initial condition for $\eqref{masterF}$ reads:
\begin{equation}
 F_n(0)=\Pr(T>0,N(0)=n)=\Pr(N(0)=n) .
\end{equation}
The random variable $N(0)$ must be understood as: ``the state of the Markov process given that an emigration event just occurred'', which is the same as :
\begin{equation}
 \Pr(N(0)=n)=K \gamma (n+1) \Pr(N(t)=n+1) ,
 \label{condini}
\end{equation}
where $K$ is a normalization constant that will be determined later. 

Summing all equations in $\eqref{masterF}$ simplifies the system considerably since almost all terms cancel out. Only the $\gamma$ terms are remaining:
\begin{equation}
\sum_{n=0}^\infty F'_n(t) = \sum_{n=0}^\infty -\gamma n  F_n(t).
  \label{formulepdf}\end{equation}
Noting $\mathit{f}_T$ the pdf of $T$, $\eqref{formulepdf}$ gives the simple relationship :
\begin{equation}
\mathit{f}_T(t)=-F'(t)=\gamma \left\langle  F_n(x)\right\rangle_n ,
  \label{formulepdf2}\end{equation}
where the brackets represent averaging. 

The general solution of $\eqref{masterF}$ for all $n$ can be obtained with the help of the generating function
\begin{equation}
G(z,t)=\sum_{n=0}^{\infty}  F_n(t) z^n ,
\label{gene}
\end{equation}
with $z\in [0,1]$. Introducing $\eqref{gene}$ into $\eqref{masterF}$ yields the following partial differential equation:
\begin{equation}
\frac{\partial G}{\partial t}-(\sigma+\mu z^2 - z (\alpha+\mu))\frac{\partial G}{\partial z}=(z-1)\lambda G  ,
\label{PDEG}
\end{equation}
where the short hand notation $\alpha=\gamma+\sigma$ is used. Equation $\eqref{PDEG}$ can be solved by means of characteristic curves. The initial condition for $\eqref{PDEG}$ is the generating function  corresponding to $\eqref{condini}$. The latter is easily obtained by observing that it is the derivative of the steady state generating function of $N$, given in \citet{ANCEY2008}:
\begin{equation}
G(z,0)=K \gamma \frac{\partial}{\partial z}\left[\left(\frac{\alpha-\mu}{\alpha-\mu z}\right)^{\lambda/\mu}\right] =\left(\frac{\alpha-\mu}{\alpha-\mu z}\right)^{\lambda/\mu+1}\nonumber ,
\label{condiniG}
\end{equation}
where $K$ have been replaced to fulfill the normalization condition $G(1,0)=1$.
The general expression of $G(z,t)$ reads:
\begin{eqnarray}
\lefteqn{G(z,t)=\left[ (z_1-z_2)e^{-\mu(1-z_2)t}\right]^{\lambda/\mu}\left(\frac{\alpha-\mu}{A(t)-B(t) z}\right)^{\lambda/\mu+1}\times}  \nonumber \\
&& \mbox{} \left[z_1-z+e^{-\mu(z_1-z_2)t}(z-z_2)\right]   ,
\end{eqnarray}
where $A(t)$, $B(t)$, $z_1$, and $z_2$ are defined as:
\begin{eqnarray}
 A(t)&=&z_2 (\mu z_1 -\alpha)e^{-\mu(z_1-z_2)t}+z_1(\alpha-\mu z_2) \nonumber \\
 B(t)&=& (\mu z_1 -\alpha)e^{-\mu(z_1-z_2)t}+(\alpha-\mu z_2)  \\
 z_1&=&(\alpha+\mu)\left(1+\sqrt{1-\epsilon}\right)/{2\mu} \nonumber \\
 z_2&=&(\alpha+\mu)\left(1-\sqrt{1-\epsilon}\right)/{2\mu} \nonumber ,
 \label{adef}
\end{eqnarray}
and $\epsilon=4\sigma\mu/(\alpha+\mu)^2$. As seen in $\eqref{formulepdf2}$, the pdf of $T$ is related to the first order moment of $ F_n$. A useful property of generating functions gives:
\begin{equation}
\left. \frac{\partial G(z,t)}{\partial z}\right|_{z=1}=\left\langle F_n\right\rangle ,
\end{equation}
so that finally:
\begin{eqnarray}
\mathit{f}_T(t)&=&\gamma \left(z_1-z_2\right)^{\lambda/\mu}\left(\frac{\alpha-\mu}{A(t)-B(t)}\right)^{\lambda/\mu+1}e^{-\lambda(1-z_2)t} \times \nonumber  \\
   \mbox{} &&\left\lbrace\frac{\left(\lambda/\mu+1\right)B(t)}{A(t)-B(t)}\left[(1-z_2)e^{-\mu(z_1-z_2)t}+z_1-1\right]\right. \nonumber\\
 &&\left. \mbox{             }\vphantom{\frac{\left(\lambda/\mu+1\right)B(t)}{A(t)-B(t)}}+ e^{-\mu(z_1-z_2)t}-1  \right\rbrace \mbox{~~~~} \forall t>0   ,
\label{pdflagtime}
\end{eqnarray}
where $A(t)$, $B(t)$, $ z_1$ and $z_2$ have been defined in $\eqref{adef}$. Equation $\eqref{pdflagtime}$ provides a non trivial and fully theoretical link between the Markov model defined on an observation window and the experimental pdf obtained at a precise location. 

\begin{figure*}[t]
\centering
\includegraphics[width=40pc]{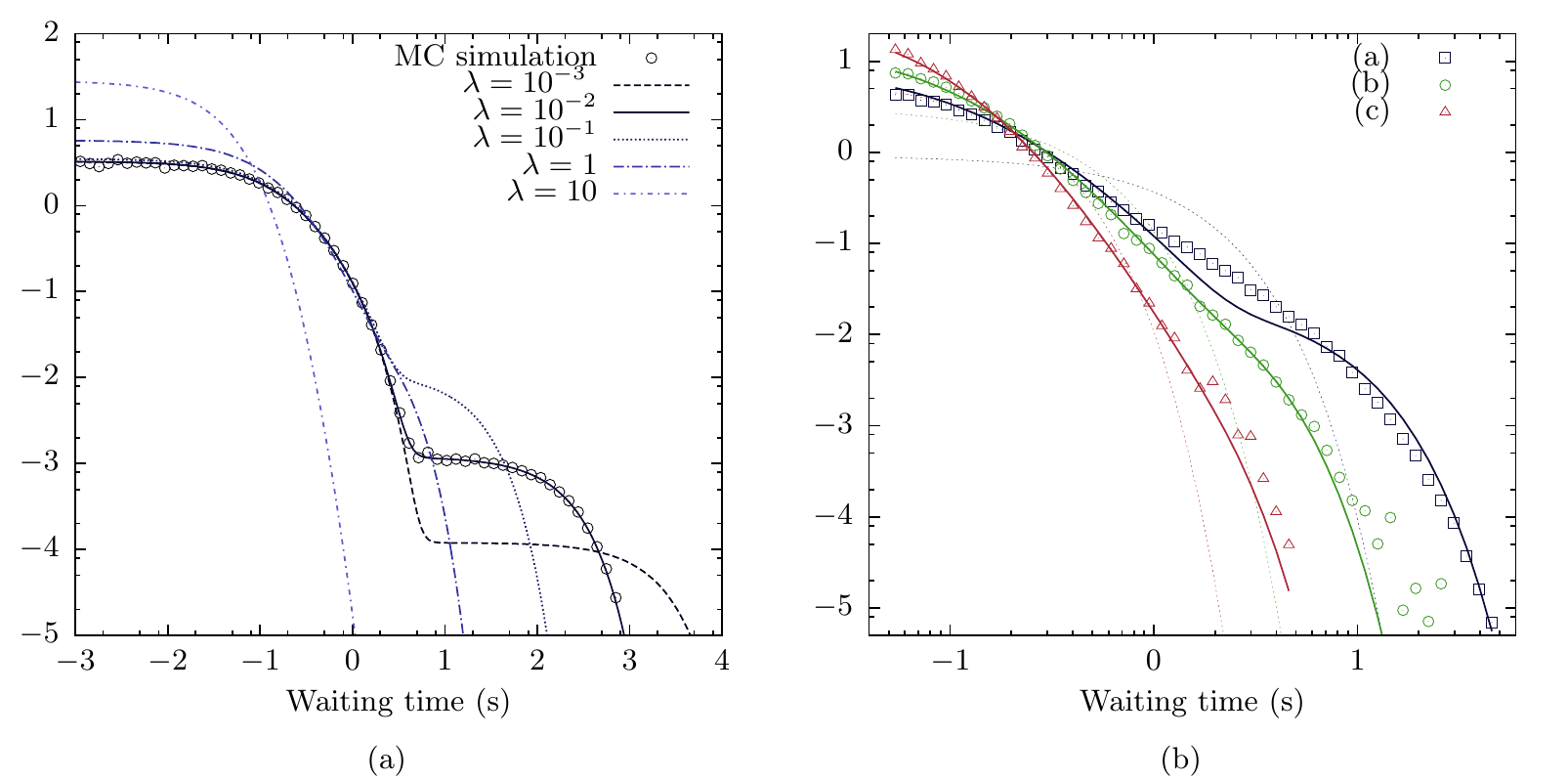}
\caption{Theoretical pdf (log-log) for the waiting time between emigration events. Constant parameters are $\sigma=1.0 \mbox{~s}^{-1}$ and $\gamma=0.5 \mbox{~s}^{-1}$. (a) Effect of the erosion rate $\lambda$ (in $\mbox{~particles.s}^{-1}$) with $\mu=1.3 \mbox{~s}^{-1}$. Comparison with a Monte Carlo simulation (for $\lambda=10^{-2}$). (b) Comparison between experimental runs (points) and theoretical pdf (lines) in log-log scale with parameters: $\lambda_{(a)}=0.56$,  $\lambda_{(b)}=2.10$ and  $\lambda_{(c)}=5.75 \mbox{~particles.s}^{-1}$. Fixed parameters are $\mu=2.37\mbox{~s}^{-1}$, $\sigma=2.2 \mbox{~s}^{-1}$ and $\gamma=0.3 \mbox{~s}^{-1}$. Exponential fits (dashed lines) are also plotted for comparison.}
\label{figure3}
\end{figure*}

To check the validity of $\eqref{pdflagtime}$, we performed Monte Carlo simulations of the Markov process. This was achieved with the Stochastic Simulation Algorithm described in \citet{Gillespie1977}. Simulations show perfect agreement with equation $\eqref{pdflagtime}$ as can be seen in Fig.~\ref{figure3}a.   As seen on the same figure, the theoretical pdf of $T$ shows, under certain conditions, a bimodal shape that differs considerably from the exponential. This bimodal distribution comes from the separation of time scales between the slow entrainment of particles and the fast collective bursts.  

To clarify this, we plotted in Fig.~\ref{figure3}a the effect of the parameter $\lambda$ (e.g., entrainment rate) on the pdf shape. When $\lambda \rightarrow 0$, with other parameters  held constant,  the time scales clearly separate. On the contrary, when $\lambda$ increases, and becomes greater than or equal to $\mu$, no further significant scale separation can be seen and the resulting pdf becomes exponential. This condition on $\lambda$  can be interpreted as the need for a slow external driving of the process for observation of a separation of time scales.

Collective entrainment is a necessary condition to observe a separation of time scales in the pdf of $T$. Indeed, even if the emigration rate is large compared to the entrainment rate and thus belongs to two distinct time scales (the fast motion of particles against the slow entrainment process), no separation is observed on the pdf when $\mu \rightarrow 0$. This can be explained by the fact that the emigration process is governed entirely by the slow entrainment (no particle leaves if no particle moves). In the case $\mu=0$, it is possible to demonstrate that the pdf is exponential with parameter $ \alpha/(\lambda \gamma) $, in agreement with \citet{Einstein}. However, when collective entrainment is considered ($\mu>0$), the separation of time scales is possible due to the occurrence of fast and intense collective bursts that occur in addition to the slow entrainment of particles by the fluid.

Expression $\eqref{formulepdf}$ can be used to fit the set of parameters $(\lambda,\gamma,\sigma,\mu)$ to the experimental pdf (Fig.~\ref{figure3}b). The value of $\lambda$ governs the long time scales and a decrease in $\lambda$ shifts the pdf tail to the right. The $\alpha-\mu$ value controls the short time scales, so that a decrease shifts the left plateau of the pdf to shorter times. 

We kept the parameters $\gamma$, $\sigma$ and $\mu$ constant for all experiments assuming that only the entrainment rate $\lambda$ increased significantly with the shear stress. This is justified by the fact that, according to traditional bedload formulas, $\lambda$ should be proportional to $\tau_s$, while $\gamma$ to $\sqrt{\tau_s}$ only. Moreover, $\sigma$ is only proportional to the falling velocity of particles while $\mu$ is almost constant \citep{ANCEY2008}. Fitted values of $\lambda$ are given in  Fig.~\ref{figure3}b. Good agreement is found for each of the three experimental pdf, both short- and long-time scales being well described along with the first order moment of $T$.

As predicted by the theory, when the Shields stress increases (and thus also the entrainment rate $\lambda$),  the experimental pdf tends to the exponential distribution. Physically speaking, the collective entrainment effect is hidden by the fast entrainment of particles. The ratio $\mbox{Var}[T]/\bar{T}$  is thus strongly reduced for high Shields numbers (Table \ref{table1}).

It is also important to point out that the present Markov model adds only one new mechanism (the collective entrainment with rate $\mu$) to the classical stochastic models, yet it predicts a much better pdf for the waiting time between emigration events compared with the common exponential distribution.

In addition to that, the model accurately described the bedload statistics in both low and moderate Shields conditions and is able to capture the transition to the exponential at higher Shields numbers. 

\section{Conclusion}
In this study we provide new insight into the collective dynamics of bedload transport down steep slope streams. The common assumption of exponential waiting time between emigration events is proved to be inadequate for bedload transport over steep slope when the flow is close to incipient motion. Indeed, the experimental pdf shows a larger relative occurrence of extremely fast or slow events.  A new explicit pdf for the waiting time is theoretically derived from a Markov model including collective entrainment and shows great similarity to the experimental pdf.

The existence of two separate time scales in the particle emigration process is observed both experimentally and in the Markov model. In the latter, the slow-time scale correspond to the entrainment of single particles by the turbulent fluid while the collective entrainment process (no matter its physical origins) is responsible for the occurrence of fast-time scale. Indeed, collective entrainment produces fast and intense bursts in the solid discharge, which greatly contribute to the large fluctuations observed.

The separation of time scales disappears for higher flow conditions (higher entrainment rates), so that an exponential waiting time is recovered at high Shields number. Although it still exists, collective entrainment is no longer distinguishable from normal entrainment. The overall transport process thus appears to fluctuate less at high transport rates than at the onset of motion. 

Finally, these results imply that measurements of sediment flux in steep slope streams, when taken at a single location (with sediment baskets, weighting traps or geophones for instance) can lead to non-exponential time statistics. This can be avoided by measuring the number of moving particles in a bed area (or particle activity as recently described by \citet{Furbish2012}) to recover the usual memoryless Markovian properties, but at the expense of much heavier instrumentation. Rather than being memoryless and independent, the emigration process alone is subjected to ageing and thus strongly depends on the starting time of the acquisition and the duration of the sample. 

\begin{acknowledgments}
This work was supported by the Swiss National Science Foundation under grant number 200021\_ 129538 (a project called ``The Stochastic Torrent: stochastic model for bedload transport on steep slope'') and by the competence center in Mobile Information and Communication Systems (grant number 5005-67322, MICS project). The authors are thankful to Belinda Bates and Jonathan Austin for their comments.
\end{acknowledgments}

\end{article}

\end{document}